\documentclass[12pt]{article}

\usepackage{amsmath,amsfonts,amssymb,fullpage,array,epsfig}

\newcommand{\bn}{\bf n}
\newcommand{\bu}{\bf u}
\newcommand{\cs}{\mathcal{S}}
\newcommand{\cd}{\mathcal{D}}
\newcommand{\ck}{\mathcal{K}}

\numberwithin{equation}{section}

\begin{document}

\begin{center}
{\LARGE The asymptotic regimes \\[2mm] of tilted
Bianchi II cosmologies}\\[4mm]
{\large C.G. Hewitt\footnote{Department of Applied
Mathematics, University of Waterloo, Waterloo, Ontario,
Canada N2L 3G1}, R. Bridson$^{1}$ and J.
Wainwright$^{1}$}
\end{center}

\vspace*{4in}
\noindent
{\small Author designated to receive
correspondence: \\
Professor John Wainwright\\ Department
of Applied Mathematics\\ University of Waterloo\\
Waterloo, Ontario, Canada N2L 3G1\\
Phone: (519) 888-4567, ext. 3623\\
Fax: (519) 746-4319\\
email: jwainwright@math.uwaterloo.ca}\\[.5in]
Suggested running head:  The asymptotic regimes 
of tilted Bianchi II cosmologies

\thispagestyle{empty}

\newpage

\begin{abstract}
In this paper we give, for the first time, a complete
description of the dynamics of tilted spatially
homogeneous cosmologies of Bianchi type II.  The source
is assumed to be a perfect fluid with equation of state
$p = (\gamma -1) \mu$, where $\gamma$ is a constant.  We
show that unless the perfect fluid is stiff, the tilt
destabilizes the Kasner solutions, leading to a
Mixmaster-like initial singularity, with the tilt being
dynamically significant.  At late times the tilt
becomes dynamically negligible unless the equation of
state parameter satisfies
$\gamma > \frac{10}{7}$.  We also find that the tilt
does not destabilize the flat FL model, with the result
that the presence of tilt increases the likelihood of
intermediate isotropization.
\end{abstract}


\section{Introduction}

Spatially homogeneous (SH) cosmologies, that is,
cosmological solutions of the Einstein field equations
that admit a local group of isometries acting on
spacelike hypersurfaces, are of considerable importance
in theoretical cosmology and have been much studied since
the 1960s.  These models can be used to analyze aspects
of the physical Universe which pertain to or which may be
affected by anisotropy in the rate of expansion, for
example, the cosmic microwave background radiation,
nucleosynthesis in the early Universe, and the question
of the isotropization of the universe itself (see, for
example, [1]).

Spatially homogeneous cosmologies also play an important
role in attempts to understand the structure and
properties of the space of all cosmological solutions of
the Einstein field equations, since they are part of a
symmetry-based hierarchy of cosmological models of
increasing complexity, starting with the familiar
 Friedmann-Lem\^{a}itre models:
\begin{itemize}
\item[i)]
Friedmann-Lem\^{a}itre cosmologies

\item[ii)]
non-tilted SH cosmologies

\item[iii)]
tilted SH cosmologies

\item[iv)]
$G_{2}$ cosmologies

\item[v)]
$G_{1}$ cosmologies

\item[vi)]
generic cosmologies
\end{itemize}

\noindent
The terminology used in this hierarchy has the following
meaning.  A SH cosmology is said to be {\it tilted}
if the fluid velocity vector is not orthogonal to the
group orbits, otherwise the model is said to be {\it
non-tilted} [2].  A $G_{2}$ cosmology
admits a local two-parameter Abelian group of isometries
with spacelike orbits, permitting one degree of freedom
as regards spatial inhomogeneity, while a $G_{1}$
cosmology admits one spacelike Killing vector field.

An important mathematical link between the various
classes in the hierarchy is provided by the idea of
representing the evolution using a {\it state space}. 
The physical state of a cosmological model at an instant
of time is represented by a point in the state space,
which is finite dimensional for classes i)-iii) and
infinite dimensional otherwise.  The Einstein field
equations are formulated as first order evolution
equations, and {\it the evolution of a cosmological model
is represented by an orbit} (i.e., a solution curve)
{\it of the evolution equations in the state space}.  The
state space of a particular class in the hierarchy is
contained in the state spaces of the more general
classes, which implies that the particular models are
represented as special cases of the more general
models.  This structure opens the possibility that the
evolution of a model in one class may be approximated,
over some time interval, by a model in a more special
class.

The models in each level of the hierarchy can be
classified according to generality in various ways.  For
our purposes the most important of these is the
algebraic classification of the isometry group of the
SH models, the so-called {\it Bianchi
classification}\footnote{This classification applies to
SH models which admit an isometry group $G_{3}$ acting
simply transitively on the hypersurfaces of homogeneity,
and includes all SH models except for the
Kantowski-Sachs models, which admit an isometry group
$G_{4}$ acting on the hypersurfaces of homogeneity, but
with no simply transitive subgroup $G_{3}$.} (see, for
example, [3], page 112).  There is also a
classification of the
$G_{2}$ cosmologies, determined by the action of the
isometry group, that is relevant for this paper.  We
refer to [4] (Table 12.4 on page
268) for details of how this classification relates to
the Bianchi classification.

In this paper we assume that the matter content of the
universe is a perfect fluid with equation of state $p =
(\gamma -1) \mu$, where $\gamma$ is constant, the cases
$\gamma =1$ (dust) and $\gamma = \frac{4}{3}$
(radiation) being of primary interest.  Considerable
work has been done in analyzing SH models,
subject to this assumption, and a detailed, although
still incomplete, description of the non-tilted models
has been obtained. We refer to  [4] 
(Chapters 6 \& 7)and [5], for details.  Much
less is known about the tilted models, but it is evident
that as one moves through stages i)-iii) in the hierarchy
there is an increase in dynamical complexity and new
features emerge at each stage.  It is plausible to
assume that as one moves to levels iv)-vi), which
contain inhomogeneous models, this trend will continue. 
Clearly, understanding the dynamics at one level of
complexity is a prerequisite for understanding the
dynamics at a higher level.  It is within this
framework, which has a long term goal of obtaining
qualitative information about the evolution of spatially
inhomogeneous models, that the analysis of SH models
assumes renewed importance.

For the class of tilted SH models, the Einstein
field equations have been written as an autonomous DE in
a number of different ways (see for example, [6], [7] 
and [8]).  Nevertheless, due to the
complexity of the equations, a detailed analysis of the
dynamics has not been given except in the case of a
subclass of models of Bianchi type V  [9]-[11].  Our
goal in this paper is to give a qualitative analysis of
the dynamics of the tilted SH  cosmologies of Bianchi
type II near the initial singularity and at late times,
using the methods of the theory of dynamical systems. 
The Bianchi II cosmologies, while very special within
the whole Bianchi class, nevertheless play a central
role since the Bianchi II state space is part of the
boundary of the state space for all higher Bianchi types
(i.e. all types except for I and V).  We thus expect
that an analysis of the dynamics of the tilted Bianchi
II class will give insight into the dynamics of the more
general tilted Bianchi classes, while providing a lower
bound for their dynamical complexity.

The paper is organized as follows.  In Section 2 we
present the evolution and constraint equations that arise
from the Einstein field equations, and in Section 3 we
describe the stability properties of the equilibrium
points of the evolution equations, which provide the
basis for the discussion of the dynamics in the singular
asymptotic regime in Section 4 and in the late time
asymptotic regime in Section 5.  In Section 6 we discuss
the implications of the results.  Appendix A gives
details of the derivation of the evolution equations.

Parts of the paper, in particular Sections 2 and 3, and
the Appendices, are inevitably of a rather mathematical
nature.  Sections 4 and 5 are less technical, and in
these sections we give the physical interpretations of
the results.  In the Introduction and in the Discussion
we discuss the longer term goals of the research and its
potential significance in a broader context.

The background material needed for this paper can be
found in [4].  In  Appendix A it is
assumed that the reader is familiar with the orthonormal
frame formalism of Ellis \& MacCallum [12] (see [4],
Chapter 1).  In addition, familiarity with some basic
concepts and results from the theory of dynamical
systems is assumed in Sections 2 and 3 (see [4], Chapter
4).  We use geometrized units with $c=1$, $8 \pi G =1$,
and the sign conventions of [4].

\section{Properties of the evolution equations}

In order to write the Einstein field equations in a form
amenable to dynamical systems analysis, we use the
orthonormal frame formalism of Ellis \& MacCallum [12]. 
In this formalism the commutation functions of the
orthonormal frame are used as the gravitational field
variables, which has the advantage of leading directly
to first order evolution equations for the gravitational
field.  The first step is to choose the orthonormal
frame to be invariant under the group of isometries,
which implies that the commutation functions depend only
on a preferred time variable $t$.  The second step is to
choose the timelike frame vector ${\bf e}_{0}$ to be
equal to the unit normal $\bn$ of the group orbits,
which is thus tangent to an irrotational congruence of
geodesics.  The third and key step is to make the
commutation functions dimensionless by dividing them by
the rate of expansion of the normal congruence, which
leads to variables that remain bounded throughout the
evolution of the models.

As described in   Appendix A, the class of tilted
Bianchi II cosmologies  can be described by the following
set of expansion-normalized variables:
\begin{equation}
{\bf x} = (\Sigma_{+}, \Sigma_{-}, \Sigma_{1},
\Sigma_{3}, N_{1}, v_{3}), \label{e2.1}
\end{equation}
subject to one constraint of the form
\begin{equation}
g({\bf x}) = h(v_{3}) \Omega -
\Sigma_{3} N_{1} = 0, \label{e2.2}
\end{equation}
where $h (v_{3})$ is given by \eqref{a.19}, i.e. 
$$
 h (v_{3}) = \frac{\sqrt{3} \gamma
v_{3}}{G}, \qquad G = 1 + (\gamma -1) v_{3}^{2}.
$$
Here $\Sigma_{+},
\Sigma_{-}, \Sigma_{1}, \Sigma_{3}$ are {\it shear
variables}, $N_{1}$ is a {\it spatial curvature
variable} and $v_{3}$ is a {\it tilt variable}.  These
variables determine the density parameter $\Omega$ and
the shear parameter $\Sigma$ according to \eqref{a.17},
\eqref{a.38} and \eqref{a.39}.  As mentioned at the end
of  Appendix A, we regard the off-diagonal shear
variables  $\Sigma_{3}$  and $\Sigma_{1}$ as representing
the two tilt degrees of freedom.

The evolution equations for the variables \eqref{e2.1},
 as derived in   Appendix A, are
given below:
\begin{equation}
\begin{split}
\Sigma_{+}' &= - (2-q) \Sigma_{+} - 3 \Sigma_{3}^{2} +
\tfrac{1}{3} N_{1}^{2} + \tfrac{1}{2 \sqrt{3}} \Sigma_{3}
N_{1} v_{3} \\
\Sigma_{-}' &= - (2-q) \Sigma_{-} + 2 \sqrt{3}
\Sigma_{1}^{2}- \sqrt{3} \Sigma_{3}^{2} - \tfrac{1}{2}
\Sigma_{3} N_{1} v_{3}
\\
\Sigma_{1}' &= - (2-q + 2 \sqrt{3} \Sigma_{-})
\Sigma_{1}\\
\Sigma_{3}' &= - (2-q - 3 \Sigma_{+} - \sqrt{3}
\Sigma_{-}) \Sigma_{3} \\
N_{1}' &= (q-4 \Sigma_{+}) N_{1} \\
v_{3}' &= \frac{v_{3} (1-v_{3}^{2})}{1- (\gamma -1)
v_{3}^{2}} (3 \gamma -4 - \Sigma_{+} + \sqrt{3}
\Sigma_{-}), 
\end{split} \label{e2.3}
\end{equation}
where
$$
q = 2 \left( 1 - \tfrac{1}{12}
N_{1}^{2} \right) - \tfrac{1}{2} G^{-1} \Omega \left[ 3
( 2 - \gamma) ( 1 - v_{3}^{2}) + 2 \gamma v_{3}^{2}
\right].
$$
  The auxiliary equation  for
$\Omega'$ is
\begin{equation}
\Omega' = G^{-1} [2 Gq - (3 \gamma -2) - (2-\gamma )
v_{3}^{2} - \gamma (\Sigma_{+} - \sqrt{3} \Sigma_{-})
v_{3}^{2} ] \Omega .\label{e2.4}
\end{equation}

The state space is the subset of $\mathbb{R}^{6}$
defined by the constraint \eqref{e2.2} and the
inequality $\Omega \geq 0$, which by \eqref{a.17},
\eqref{a.38} and \eqref{a.39}, is equivalent to
\begin{equation}
\Sigma_{+}^{2} + \Sigma_{-}^{2} + \Sigma_{1}^{2}
+\Sigma_{3}^{2} + \tfrac{1}{12} N_{1}^{2} \leq 1.
\label{e2.5}
\end{equation}
This restriction, and the fact that $v_{3}^{2}< 1$,
implies that the state space is bounded.

The evolution equations \eqref{e2.3} are invariant under
the transformations
$$
(\Sigma_{+}, \Sigma_{-}, \Sigma_{1}, \Sigma_{3}, N_{1}, 
v_{3}) \rightarrow (  \Sigma_{+}, \Sigma_{-}, \pm
\Sigma_{1}, \pm \Sigma_{3}, \pm N_{1}, \pm v_{3}),
$$
provided that the product $v_{3} \Sigma_{3} N_{1}$ does
not change sign.  In addition, it follows that 
$N_{1}, \Sigma_{1}, \Sigma_{3}$ and $v_{3}$ cannot
change sign along an orbit.  Thus, without loss of
generality, we can assume that
\begin{equation*}
N_{1} \geq 0, \quad \Sigma_{1} \geq 0, \quad \Sigma_{3}
\geq 0, \quad \text{and} \quad v_{3} \geq 0. 
\end{equation*}
Taking these restrictions into account, the state space
$\cd$ of the tilted perfect fluid Bianchi II cosmologies
is defined by the inequalities
\begin{equation}
N_{1} > 0, \quad \Omega >0, \quad 0 < v_{3} < 1, \quad
\Sigma_{3} >0, \quad \Sigma_{1} \geq 0. \label{e2.6}
\end{equation}
The boundary $\partial \cd$ is obtained by
successively replacing the strict inequalities in
\eqref{e2.6} by equalities.

The evolution equations \eqref{e2.3} are an autonomous
DE in $\mathbb{R}^{6}$ of the form
$$
{\bf x}' = {\bf f} ({\bf x}),
$$
where the function ${\bf f}: \mathbb{R}^{6} \rightarrow
\mathbb{R}^{6}$ on the right side is a {\it rational
function} (note the function $G$ in the denominator in
$q$, and the form of the $v_{3}^{\prime}$
equation).  For values of $\gamma$ that satisfy $0 <
\gamma < 2$, the two functions in the denominator are
strictly positive on the physical state space $\cd$ and
on its boundary $\partial \cd$.  The DE \eqref{e2.3} is
thus smooth, indeed analytic, on the set $\cd \cup
\partial \cd$.  Since this set is compact and invariant,
the solutions of the DE \eqref{e2.3} can be extended for
all $\tau \in \mathbb{R}$.

The constraint \eqref{e2.2} entails a consistency
requirement, namely that the equation $g = 0$ should
define an invariant set of the evolution equations
\eqref{e2.3}.  A straightforward calculation using
\eqref{e2.2}-\eqref{e2.4} shows that
\begin{equation}
g' = (2 q -2 - \Sigma_{+} + \sqrt{3} \Sigma_{-} ) g.
\label{e2.7}
\end{equation}
It thus follows (see [4], Proposition 4.1 on page 29)
that $g =0$ does indeed define an invariant set.

In analyzing a class of Bianchi cosmologies, one
typically finds that the orbits in the {\it boundary of
the state space} play a significant role in determining
the dynamics, since orbits in the interior can
shadow\footnote{i.e. approximate closely.  See [4], page
104.} orbits in the boundary.  For the tilted Bianchi II
models, the boundary of the state space is the union of
five  disjoint invariant sets, as shown in Table 2.1.  The invariant sets i)-iv) describe familiar solutions
but give multiple representations of them.  For example,
the orbits with $\Sigma_{1} = 0$ in the invariant set
i) describe non-tilted Bianchi II cosmologies relative
to a Fermi-propagated frame, while the orbits with
$\Sigma_{1} >0$ describe the same models, but relative
to a rotating frame (see \eqref{a.37}).  Likewise, the
orbits with
$\Sigma_{1} =
\Sigma_{3} =0$ in the invariant set ii) describe
(nontilted) Bianchi I cosmologies relative to a
Fermi-propagated frame, while the orbits with
$\Sigma_{1} >0$ and/or $\Sigma_{3} >0$ describe the same
models, but relative to a rotating frame.

\begin{center}
\begin{tabular}{lccc}
& {\bf Name} & {\bf Restrictions} & {\bf Dimension} \\
\hline
& & \\
i) &non-tilted non-vacuum Bianchi II & $v_{3} = 0 =
\Sigma_{3}$, $N_{1} > 0$, $\Omega > 0$ & 4 \\[3mm]
ii) & non-tilted non-vacuum Bianchi I & $v_{3} = 0 =
N_{1}$,
$\Omega > 0$ & 4 \\[3mm]
iii) & Taub (vacuum Bianchi II) & $\Sigma_{3} = 0 =
\Omega$,
$N_{1} >0$, $v_{3} < 1$ & 4 \\[3mm]
iv) & Kasner (vacuum Bianchi I) & $N_{1} = 0 = \Omega$,
$v_{3} < 1$ & 4 \\[3mm]
v) & extreme tilt set $(\gamma < 2)$ & $v_{3} = 1$ &
4\\[2mm]
\hline
\end{tabular}

\bigskip
Table I: The invariant sets that comprise the boundary
of the state space \eqref{e2.6}.
\end{center}

The invariant set iii) contains the usual representation
of the Taub vacuum solutions ([4] pages 137-8 and page
196) when $\Sigma_{1} = 0 = v_{3}$, but it also contains
multiple representations of these solutions relative to
a non-Fermi-propagated  frame.  Likewise, the
invariant set iv) contains the usual representation of
the  well-known Kasner vacuum solutions when $\Sigma_{1}
= \Sigma_{3} = v_{3} = 0$, but it also contains multiple
representations of these solutions.  The orbits in the
remaining part of the boundary, the extreme tilt set v),
do not correspond to spatially homogeneous cosmological
solutions of the Einstein field equations since equation
\eqref{a.21} breaks down when $v_{3} \rightarrow 1$ (i.e.
$v_{b} v^{b} \rightarrow 1$).

In addition to the above invariant sets, the DE
\eqref{e2.3} has one invariant set that is not a part of
the boundary, namely the invariant set defined by
$\Sigma_{1} =0$.  In view of the interpretation of
$\Sigma_{1}$, this invariant set corresponds to tilted
Bianchi II models with one tilt degree of freedom.  We
shall see that the dynamics of these models is
significantly simpler than the dynamics of the full
class.

\bigskip
\noindent
{\it Technical point}:

\medskip

It is necessary for the subsequent analysis to determine
at which points the constraint surface \eqref{e2.2} is
singular (i.e. the points ${\bf x} \in \mathbb{R}^{6}$
which satisfy $g ({\bf x}) = 0$ and $\nabla g ({\bf x})
= \pmb{0})$, since at these points one cannot use the
implicit function theorem to eliminate, locally, one of
the variables.  The gradient of $g$ is
\begin{equation}
\nabla g = \left( -2h \Sigma_{+}, -2 h \Sigma_{-}, -2h
\Sigma_{1}, -2h \Sigma_{3} - N_{1}, - \tfrac{1}{6} h
N_{1} - \Sigma_{3}, h' \Omega \right) , \label{e2.8}
\end{equation}
where the variables are listed in the order \eqref{e2.1}.

It follows that the surface is singular at and only at
points ${\bf x}$ given by $v_{3} = N_{1} = \Sigma_{3} =
\Omega = 0$.  Thus the surface is non-singular at all
points of the state space \eqref{e2.6}, and is singular
only on part of the boundary, i.e. a two-dimensional
subset of the Kasner set.  Indeed, if $\Omega >0$, the
constraint can be written in the form
$$
h (v_{3}) = \frac{\Sigma_{3} N_{1}}{\Omega} ,
$$
and since $h' (v_{3}) >0$ for $0 \leq v_{3} \leq 1$ and
$0 < \gamma < 2$, it follows that $v_{3}$ is determined
uniquely in terms of the other variables.

It follows from \eqref{e2.8} that the vectors
\begin{equation}
{\bf e}_{A} = \frac{\partial}{\partial x_{A}} - \frac{1}{h'
(v_{3}) \Omega} \frac{\partial g}{\partial x_{A}}
\frac{\partial}{\partial v_{3}}, \label{e2.9}
\end{equation}
where $A = 1,2, \ldots , 5$ and $(x_{A}) = (\Sigma_{+},
\Sigma_{-}, \Sigma_{1}, \Sigma_{3}, N_{1})$, satisfy
${\bf e}_{A} \cdot \nabla g = 0$ and are linearly
independent.  These vectors thus span the tangent space
to the constraint surface $g = 0$ in $\mathbb{R}^{6}$. 
Equivalently, an orbit lies in the constraint surface $g
=0$ if and only if its tangent vector is a linear
combination of the
${\bf e}_{A}$.

\section{Equilibrium points}

In this Section we consider the local stability of the
equilibrium points of the DE \eqref{e2.3} and the
constraint \eqref{e2.2}, i.e. points ${\bf x} = {\bf a}$
that satisfy
\begin{equation}
{\bf f} ({\bf a}) = \pmb{0}, \qquad g ({\bf a}) = 0.
\label{e3.1}
\end{equation}
The local stability is determined by linearizing the DE
\eqref{e2.3} at ${\bf x} = {\bf a}$, which gives
$$
{\bf x}' = D {\bf f} ({\bf a}) {\bf x},
$$
and finding the eigenvalues of the derivative matrix $D
{\bf f} ({\bf a})$.  The analysis is complicated by the
constraint, which requires that we consider only
eigenvectors that are tangent to the constraint surface,
i.e. that are orthogonal to the gradient vector $\nabla
g ({\bf a})$.  We shall refer to eigenvalues and
eigenvectors that satisfy this condition as
{\it physical}.  We note that if all the physical
eigenvalues have negative (positive) real parts then the
equilibrium point is a local sink (source), i.e. it
attracts (repels) all orbits in a neighbourhood.  In
addition to isolated equilibrium points, we will also
encounter arcs of equilibrium points, for which one
eigenvalue is necessarily zero (see for example [4],
Section 4.3.4).  In this case the criterion for a local
sink (source) is that all eigenvalues other than the
zero one have negative (positive) real parts.

We now list the equilibrium points, obtained by
systematically solving equations \eqref{e3.1}.  Each
equilibrium point, apart from those with extreme tilt,
corresponds to a self-similar solution of the Einstein
field equations\footnote{The situation is analogous to
the non-tilted case ([4], Section 5.2.3).}, which we also
give.

\bigskip
\noindent
{\it Non-vacuum equilibrium points $(\Omega >
0)$:}

\begin{itemize}
\item[i)]
{\it Flat FL  point, $F$}
$$
\Sigma_{+} = \Sigma_{-} = \Sigma_{1} = \Sigma_{3} =
N_{1} = v_{3} = 0,
$$
$$
\Omega = 1, \quad \Sigma = 0, \quad q = \tfrac{1}{2} (3
\gamma -2), \quad 0 < \gamma \leq 2.
$$

{\it Self-similar solution}: the flat FL solution.

\item[ii)]
{\it Non-tilted point, PII}
$$
\Sigma_{+} = \tfrac{1}{8} (3 \gamma -2), \quad N_{1} =
\tfrac{3}{4} \sqrt{(3 \gamma -2) (2- \gamma )}, \quad
\Sigma_{-} = \Sigma_{1} = \Sigma_{3} = v_{3} = 0,
$$
$$
\Omega = \tfrac{3}{16} ( 6 - \gamma ), \quad \Sigma =
\tfrac{1}{8} (3 \gamma - 2), \quad q = \tfrac{1}{2} (3
\gamma - 2), \quad \tfrac{2}{3} < \gamma < 2 .
$$

{\it Self-similar solution}: the Collins-Stewart
solutions([4], pages 131 and 189).

\item[iii)]
{\it Tilted point, PII$_{\it tilt}$}
$$
\Sigma_{+} = \tfrac{1}{8} (3 \gamma -2), \quad
\Sigma_{-} = \tfrac{\sqrt{3}}{8} (10 - 7 \gamma ), \quad
\Sigma_{3} =  \tfrac{1}{4} \alpha \sqrt{(11 \gamma -10)
(7 \gamma -10)}, \quad \Sigma_{1} = 0, 
$$
$$
N_{1} =  \alpha \sqrt{3(5 \gamma -4) (3 \gamma
-4)}, \quad v_{3} = \sqrt{\frac{(3 \gamma -4)(7 \gamma
-10)}{(11 \gamma -10)(5 \gamma -4)}}, 
$$
$$
\alpha^{2} = \frac{3 (2- \gamma )}{17 \gamma - 18},
\quad \tfrac{10}{7} < \gamma < 2.
$$
$$
\Omega = \tfrac{1}{4} \alpha^{2} (21 \gamma^{2} - 24
\gamma + 4), \quad \Sigma^{2} = 1 - \alpha^{2} (\gamma
-1) (9\gamma -5),
\quad q = \tfrac{1}{2} (3
\gamma -2).
$$

{\it Self-similar solution}: first given in [13].

\item[iv)]
{\it Line of tilted points, $\mathcal{L}$II$_{\it tilt}$}
$$
\Sigma_{+} = \tfrac{1}{3}, \quad \Sigma_{-} = -
\tfrac{1}{3 \sqrt{3}}, \quad \Sigma_{3} = \tfrac{2}{3}
\sqrt{\tfrac{1}{57} (4b+1) (8-3b)}, \quad \Sigma_{1} =
\tfrac{2}{3 \sqrt{3}} b,
$$
$$
N_{1} = 2 \sqrt{\tfrac{1}{57} (2b+1) (17-8b)}, \quad
v_{3} = \sqrt{\frac{3(4b+1) (2b+1)}{(17-8b) (8-3b)}}
$$
with
$$
0 < b < 1, \qquad \gamma = \tfrac{14}{9}.
$$
$$
\Omega = \tfrac{2}{171} (16 b^{2} - 45b + 59), \quad
\Sigma^{2} = \tfrac{4}{171} (2b+1) (9-2b), \quad q =
\tfrac{4}{3}.
$$

{\it Self-similar solutions}: not given previously.

\item[v)]
{\it Extreme tilted point, PII$_{\it extreme}$}
$$
\Sigma_{+} = \tfrac{1}{3}, \quad \Sigma_{-} = -
\tfrac{1}{3 \sqrt{3}}, \quad \Sigma_{3} = \tfrac{10}{3
\sqrt{57}}, \quad \Sigma_{1} = \tfrac{2}{3 \sqrt{3}},
\quad N_{1} = \tfrac{6}{\sqrt{19}}, \quad v_{3} =1
$$
$$
\Omega = \tfrac{20}{57}, \quad \Sigma^{2} =
\tfrac{28}{57}, \quad q = \tfrac{4}{3}, \quad 0 < \gamma
< 2.
$$

\item[vi)]
{\it Jacobs disc}, $\mathcal{J}$
$$
\Sigma_{1} = \Sigma_{3} = N_{1} = v_{3} = 0, \quad
\Sigma_{+}^{2} + \Sigma_{-}^{2} < 1
$$
$$
\Omega > 0, \quad \Sigma < 1, \quad q =2, \quad \gamma
=2.
$$

{\it Self-similar solutions}: the Jacobs stiff fluid
solutions ([14], page 1109).

\end{itemize}

\bigskip
\noindent
{\it Vacuum equilibrium points $(\Omega =0)$}:

\begin{itemize}
\item[i)]
{\it Kasner circle, $\mathcal{K}$}
$$
\Sigma_{+}^{2} + \Sigma_{-}^{2} = 1, \quad \Sigma_{3} =
\Sigma_{1} = N_{1} = v_{3} = 0
$$
$$
\Omega = 0, \quad \Sigma =1, \quad q = 2, \quad 0 <
\gamma \leq 2.
$$

{\it Self-similar solutions}: the Kasner vacuum
solutions ([4], pages 132 and 188)

\item[ii)]
{\it Kasner circle with extreme tilt, $\mathcal{K}_{\it
extreme}$}
$$
\Sigma_{+}^{2} + \Sigma_{-}^{2} =1, \quad \Sigma_{3} =
\Sigma_{1} = N_{1} = 0, \quad v_{3} =1
$$
$$
\Omega =0, \quad \Sigma =1, \quad q = 2, \quad 0 <
\gamma < 2.
$$

\item[iii)]
{\it Kasner lines with tilt, $\mathcal{K}^{\pm}_{\it
tilt}$}
$$
\Sigma_{+} = \sqrt{3} \Sigma_{-} + 3 \gamma -4, \quad
\Sigma_{-} = \tfrac{\sqrt{3}}{4} [ - 3 \gamma + 4 \pm
\sqrt{(3 \gamma -2) (2 - \gamma )} ], \quad \Sigma_{3} =
\Sigma_{1} = 0
$$
$$
N_{1} = 0, \qquad 0 < v_{3} < 1
$$
$$
\Omega = 0, \quad \Sigma =1, \quad q = 2, \quad
\tfrac{2}{3} \leq \gamma \leq 2.
$$

{\it Self-similar solutions}: the Kasner vacuum
solutions referred to a non-Fermi-propagated frame.
\end{itemize}

\bigskip

If $\gamma$ satisfies $\frac{2}{3} < \gamma
< 2$, there are two Kasner lines $\mathcal{K}_{\rm
tilt}^{\pm}$,  which join
$\mathcal{K}$ to $\mathcal{K}_{\rm extreme}$, while if
$\gamma =
\frac{2}{3}$ or $\gamma =2$, the two lines coincide,
with $(\Sigma_{+}, \Sigma_{-}) = \left( - \frac{1}{2},
\frac{\sqrt{3}}{2} \right)$ and $(\Sigma_{+},
\Sigma_{-}) = \left( \frac{1}{2}, - \frac{\sqrt{3}}{2}
\right)$, respectively.

\bigskip

We need to know whether any of the equilibrium points
are local sinks or sources.  It turns out that for each
value of $\gamma$ in the interval $0 < \gamma < 2$ one
of the equilibria, with $\Omega > 0$ is a local
sink\footnote{With the exception of $F$, many of the
eigenvalues are complicated expressions in $\gamma$;
their explicit form is unimportant for our purposes.},
as indicated in Table 3.1. The equilibria are related to
one another by a series of bifurcations that occur as
$\gamma$ varies.  By inspection of the coordinates of
these equilibrium points, we observe the following
transitions:
\begin{equation}
F \xrightarrow{\gamma = \frac{2}{3}} {\rm PII}
\xrightarrow{\gamma =
\frac{10}{7}} {\rm PII}_{\rm tilt}
\xrightarrow[b=0]{\gamma = \frac{14}{9}} \mathcal{L}{\rm
II}_{\rm tilt} \xrightarrow[b=1]{\gamma = \frac{14}{9}}
{\rm PII}_{\rm extreme} \label{e3.2}
\end{equation}

\begin{center}
\begin{tabular}{cc}
{\bf Range of $\boldsymbol{\gamma}$} & {\bf Local sink}
\\ \hline
& \\
$0 < \gamma \leq \frac{2}{3}$ & flat FL point $F$ \\[3mm]
$\frac{2}{3} < \gamma \leq \frac{10}{7}$ & non-tilted
point PII \\[3mm]
$\frac{10}{7} < \gamma < \frac{14}{9}$ & tilted point
PII$_{\rm tilt}$ \\ [3mm]
$\gamma = \frac{14}{9}$ & line of tilted points
$\mathcal{L}$II$_{\rm tilt}$ \\[3mm]
$\frac{14}{9} < \gamma < 2$ & extreme tilted point
PII$_{\rm extreme}$ \\[2mm] \hline 
\end{tabular}

\bigskip
Table II: Local sinks in the tilted Bianchi II state
space.
\end{center}

We can describe the mechanisms for these bifurcations,
without giving full details, as follows.  The
linearization of the evolution equation for $N_{1}$ at
$F$ is
$$
N_{1}' = \tfrac{1}{2} (3 \gamma - 2) N_{1},
$$
showing that the spatial curvature variable $N_{1}$
destabilizes
$F$ at
$\gamma =
\frac{2}{3}$.  The associated eigenvector is ${\bf
e}_{5}$ in \eqref{e2.9}.  The linearizations of the
evolution equations for $\Sigma_{3}$ and $v_{3}$ at PII
are
$$
\Sigma_{3}' = \tfrac{3}{8} (7 \gamma -10) \Sigma_{3},
\quad v_{3}' =\tfrac{3}{8} (7 \gamma -10) v_{3},
$$
showing that $v_{3}$ and $\Sigma_{3}$ destabilize PII at
$\gamma = \frac{10}{7}$.  The associated physical
eigenspace is actually one-dimensional and the
eigenvector is ${\bf e}_{4}$.  Finally, the
linearization of $\Sigma_{1}'$ at PII$_{\rm tilt}$ is
$$
\Sigma_{1}' = \tfrac{3}{4} (9 \gamma - 14) \Sigma_{1},
$$
showing that $\Sigma_{1}$ destabilizes PII$_{\rm tilt}$
at $\gamma = \frac{14}{9}$.  Stability is transferred
from PII$_{\rm tilt}$ to PI$_{\rm extreme}$ through the
line of equilibrium points $\mathcal{L}$II$_{\rm tilt}$,
which exists only for $\gamma = \frac{14}{9}$.  We shall
refer to the bifurcation at $\gamma = \frac{2}{3}$ as
{\it the spatial curvature bifurcation}, and, in view of
the fact that $\Sigma_{3}$ and $\Sigma_{1}$ represent
the tilt degrees freedom, we shall refer to the
bifurcations at $\gamma = \frac{10}{7}$ and $\gamma =
\frac{14}{9}$ as {\it the first and second tilt
bifurcations}.

As regards local sources, it turns out that unless
$\gamma =2$, none of the equilibrium points or
equilibrium sets is a local source.  This result follows
from a careful analysis of the eigenvalues associated
with the equilibrium points and sets.  If $\gamma =2$,
it turns out that a subset of the Jacobs disc shown as
the shaded region in Figure I, is a local source.

\begin{center}

\epsfig{file=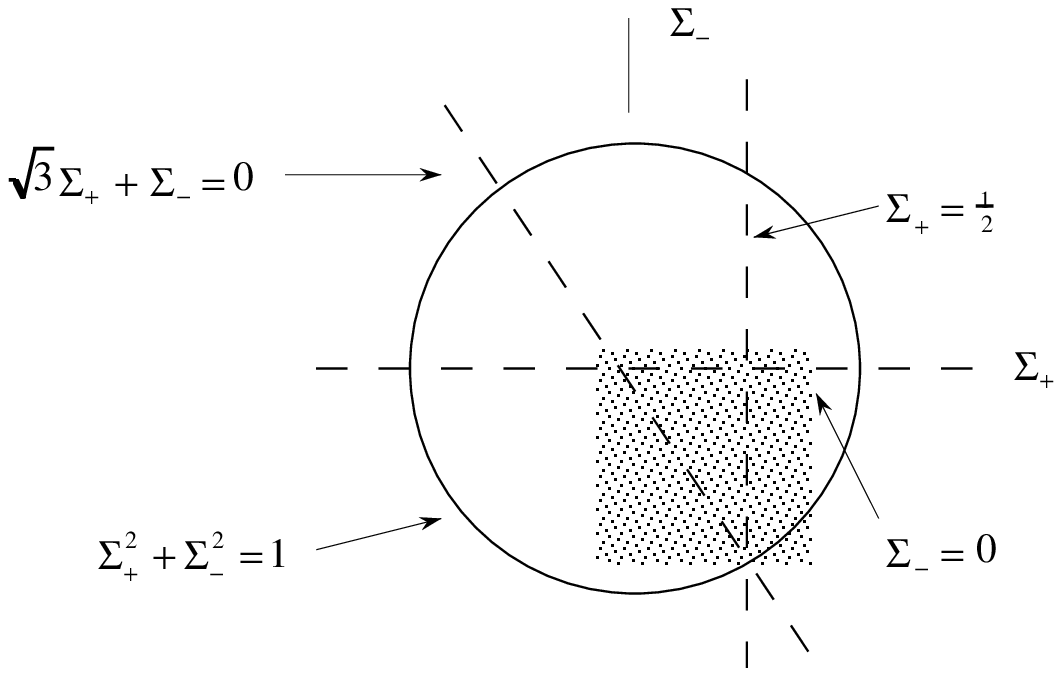,width=5in}

{\small {\bf Figure I:} The shaded region, a subset of
the Jacobs disc, is a local source in the case $\gamma
=2$.}
\end{center}

\bigskip

It is of interest to consider the invariant subset
defined by $\Sigma_{1} = 0$, which describes the
evolution of models with one tilt degree of freedom. 
For these models an analysis of the eigenvalues shows
that there is an arc of the Kasner circle $\ck$, defined
by $|\Sigma_{+}| < \frac{1}{2}$, $\Sigma_{-} > 0$, that
is a local source.

\section{The late time asymptotic regime}

We have seen that for each value of $\gamma$ in the
range $0 < \gamma < 2$, excluding $\gamma =
\frac{14}{9}$, there is a unique equilibrium point that
is a local sink of the evolution equations, while if
$\gamma = \frac{14}{9}$, there is an arc of equilibrium
points that is a local sink. These local sinks are
listed in Table 3.1, and the bifurcations that occur
as $\gamma$ increases are displayed in equation
\eqref{e3.2}.   By definition of local sink, any orbit
that enters a sufficiently small neighbourhood of the
sink approaches the sink as $\tau
\rightarrow + \infty$.  The monotone functions in Appendix B,
and numerical simulations, provide strong evidence that
{\it for a given value of $\gamma$ the local sink is the
future attractor of the evolution equation}, i.e. all
orbits, except possibly a set of measure zero, approach
the local sink as $\tau \rightarrow + \infty$.

The main conclusion that can be drawn from this
asymptotic result is that  the dynamical
significance of the tilt and the shear at late times
increases as the equation of state parameter increases
from 0 to 2, as follows:

\begin{itemize}
\item[i)]
  in the range $0 < \gamma <
\frac{2}{3}$ the models isotropize, and since the
deceleration parameter is asymptotically negative (i.e.
$\lim\limits_{\tau \rightarrow + \infty} q = \frac{1}{2} (3
\gamma -2) < 0$), the models are inflationary, 

\item[ii)] at $\gamma = \frac{2}{3}$ the spatial
curvature destabilizes the flat FL equilibrium point,
and for
$\gamma > \frac{2}{3}$ the models no longer isotropize,

\item[iii)]
at $\gamma = \frac{10}{7}$, the tilt destabilizes the
Collins-Stewart solution, and for $\gamma >
\frac{10}{7}$, the models are asymptotically tilted at
late times, and

\item[iv)]
if $\gamma > \frac{14}{9}$, the tilt is asymptotically
extreme $(v \rightarrow 1)$ at late times.
\end{itemize}

 The $\gamma$-dependent limits of the dimensionless shear
scalar $\Sigma$ and the tilt variable $v$, as defined by
\eqref{a.23}, can be obtained from the list of
equilibrium points in Section 3.  In the physically
important cases of dust $(\gamma =1)$ and radiation
$\left( \gamma = \frac{4}{3} \right)$, which satisfy
$\frac{2}{3} < \gamma < \frac{10}{7}$, the models are
asymptotic to the Collins-Stewart solution i.e. they
do not isotropize $(\Sigma \nrightarrow 0)$, but the
tilt becomes dynamically negligible $(v \rightarrow 0)$, at
late times.

\section{The singular asymptotic regime}

As mentioned in Section 3, there is no equilibrium point
or equilibrium set that is a local source, except in the
special case $\gamma =2$, or unless one restricts
consideration to models with only one tilt degree of
freedom (the invariant set $\Sigma_{1} = 0$).  The
implication of this fact is that {\it a typical orbit is
not past asymptotic to an equilibrium point}.  The
situation is analogous to the case of non-tilted SH
cosmologies of Bianchi types VIII and IX (see [4],
Section 6.4), for which it has been shown that there
exist {\it infinite heteroclinic sequences} based on the
circle
$\mathcal{K}$ of Kasner equilibrium points, i.e.
infinite sequences of equilibrium points on
$\mathcal{K}$, joined by special heteroclinic orbits
directed into the past (see [15] and [4],
Section 6.4.2, for a geometrical description of these
heteroclinic sequences). These heteroclinic sequences
determine the dynamics in the singular regime
$(\tau \rightarrow - \infty )$ in the sense that a typical
orbit shadows\footnote{In the language of dynamical
systems, we are describing the $\alpha$-limit set of a
typical orbit.  For recent progress in proving the
existence of the $\alpha$-limit set,we refer to [16] \&
[17.} (i.e. is approximated by) a
heteroclinic sequence as
$\tau
\rightarrow - \infty$.  In physical terms, the dynamics of a
typical cosmological model is approximated by a sequence
of Kasner vacuum models as the singularity is approached
into the past, the so-called {\it Mixmaster oscillatory
behaviour}.

The present situation is more complicated due to the
existence of two Kasner circles, the standard Kasner
circle $\mathcal{K}$, and the Kasner circle
$\mathcal{K}_{\rm extreme}$, with extreme tilt (i.e.
$v_{3} =1$).  The heteroclinic sequences contain orbits
that join two points on $\mathcal{K}$, orbits that join
two points on $\mathcal{K}_{\rm extreme}$, and orbits
that join a point on $\mathcal{K}$ to a point on
$\mathcal{K}_{\rm extreme}$, and vice versa.  There are
three families of orbits that join two points on the
same Kasner circle.  These orbits satisfy $\Omega =0$,
i.e.
$$
\Sigma_{+}^{2} + \Sigma_{-}^{2} + \Sigma_{1}^{2} +
\Sigma_{3}^{2} + \tfrac{1}{12} N_{1}^{2} =1,
$$
have one of $\Sigma_{1}, \Sigma_{3}$ and $N_{1}$
non-zero, and have
$$
v_{3} = 0 \quad \text{or} \quad v_{3} =1.
$$
The three families are given by

\begin{itemize}
\item[i)]
$N_{1}>0$, $\Sigma_{-} = C_{1} (\Sigma_{+} -2)$,

\item[ii)]
$\Sigma_{3} >0$, $\Sigma_{+} - \sqrt{3} \Sigma_{-} =
C_{2}$,

\item[iii)]
$\Sigma_{1} >0$, $\Sigma_{+} =C_{3}$,
\end{itemize}

\noindent
where $C_{1}, C_{2}$ and $C_{3}$ are constants.  The
projections of these orbits in the $\Sigma_{+}
\Sigma_{-}$-plane are shown in Figure II, with the
arrows showing evolution into the past.  We note that
the orbits i) describe the Taub vacuum solutions of
Bianchi type II (see [4], page 137), while the orbits ii)
and iii) describe the Kasner vacuum solutions relative
to a rotating frame (i.e. the angular velocity
$R_{\alpha}$ of the spatial frame is non-zero; see
Appendix A).

\begin{center}

\epsfig{file=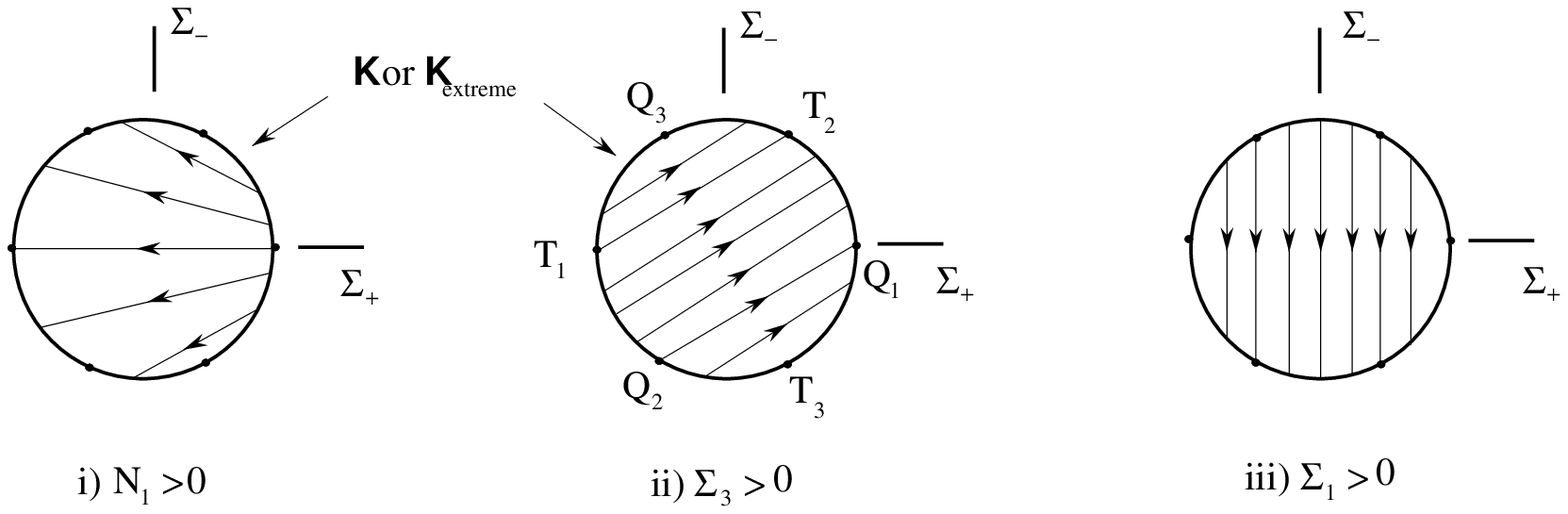,width=6.5in}

{\small{\bf Figure II:} Vacuum orbits joining points on
the Kasner circle $\ck$ or the extreme Kasner
circle $\ck_{\rm extreme}$, given by $\Sigma_{+}^{2} +
\Sigma_{-}^{2} =1$.  The arrows show evolution into the
past.}
\end{center}

\bigskip

The orbits joining an equilibrium point on $\ck$ to an
equilibrium point on $\ck_{\rm extreme}$ are given by 
\begin{itemize}
\item[iv)]
$\Sigma_{+}, \Sigma_{-} = \text{constant}$, \quad
$\Sigma_{+}^{2} + \Sigma_{-}^{2} =1$, \quad  $\Sigma_{1}
= \Sigma_{3} = N_{1} = 0$,
\end{itemize}

\noindent
with $v_{3}' < 0$ (i.e. $v_{3}$ varying between 0 and 1
for evolution into the past), or $v_{3}' >0$.  The
direction of flow along these orbits is determined by
the Kasner lines with tilt $\ck^{\pm}$, which depend on
the value of $\gamma$, and is shown in Figure III.

\begin{center}

\epsfig{file=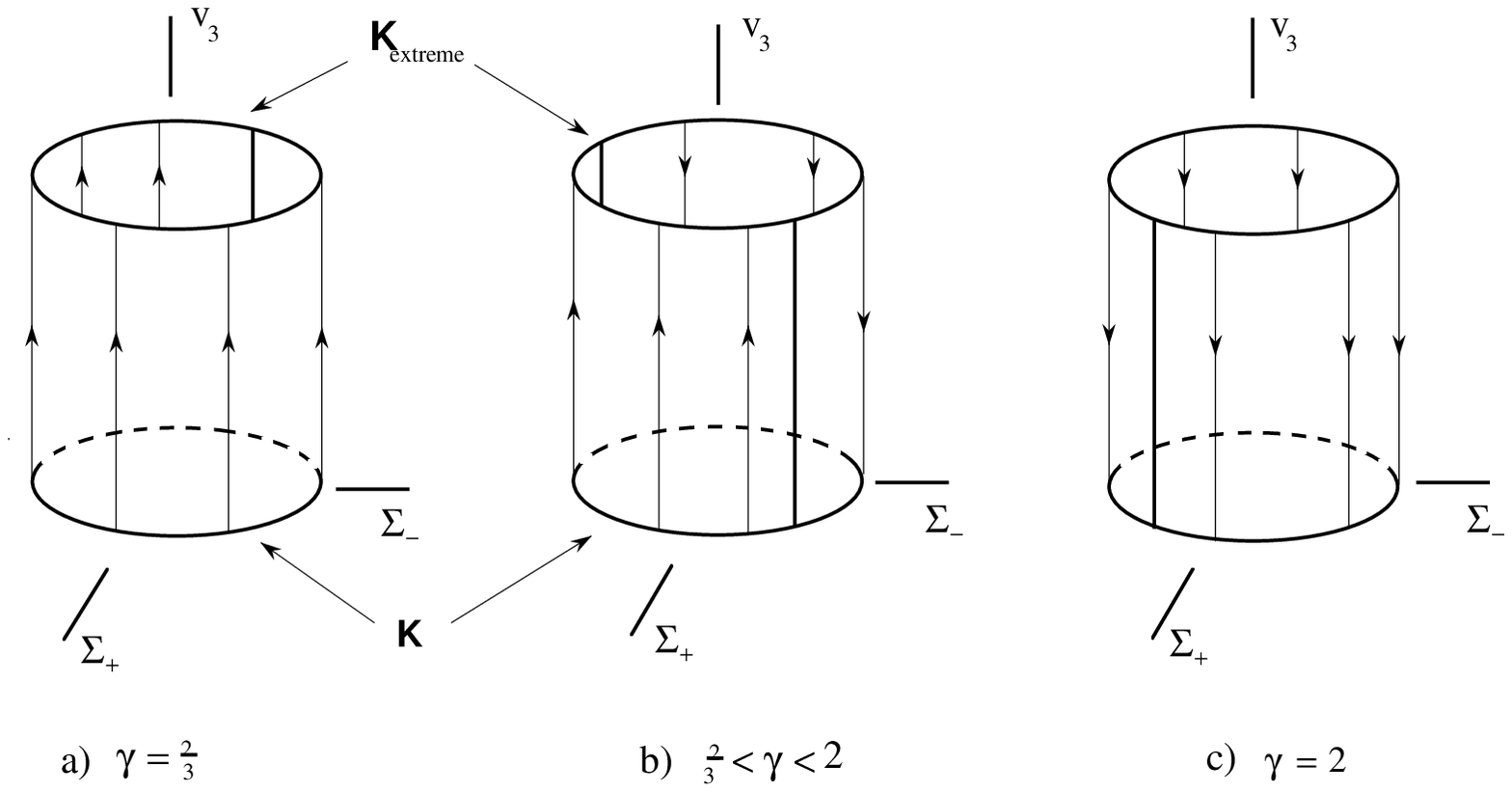,width=6.5in}

{\small{\bf Figure III:}\footnote{These invariant sets
of orbits joining $\ck$ and $\ck_{\rm extreme}$ first
appeared in the analysis of the tilted SH models of
Bianchi type V (see [10] and [11]).}  Vacuum orbits
joining points on the Kasner circle
$\ck$ and the extreme Kasner circle
$\ck_{\rm extreme}$.  The arrows show evolution into the
past.}
\end{center}

\bigskip

\noindent
If $0 < \gamma \leq \frac{2}{3}$, the flow into the past
along these orbits is from $\ck$ to $\ck_{\rm extreme}$,
while if $\gamma =2$, the flow into the past is from
$\ck_{\rm extreme}$ to $\ck$.  On the other hand, if
$\frac{2}{3} < \gamma < 2$, one family of orbits links
$\ck$ to $\ck_{\rm extreme}$ while the second family
does the reverse.

We can now describe the infinite heteroclinic
sequences.  In the case $\frac{2}{3} < \gamma < 2$, the
orbits that join successive Kasner points belong to the
eight families i)-iii) with $v_{3} =0$ or $v_{3} =1$ and
iv) with $v_{3}' >0$ or $v_{3}' < 0$.  In the case $0 <
\gamma \leq \frac{2}{3}$, since $\lim\limits_{\tau \rightarrow
- \infty} v_{3} =1$, only orbits in the three families
i)-iii) with $v_{3} =1$ are permitted.

It is not our intention to describe the detailed
structure of the heteroclinic sequences, as has been
done in the case of the non-tilted models of Bianchi
types VIII and IX (see the references at the beginning
of this section).  We simply wish to point out that
{\it the evolution of tilted cosmologies of Bianchi type
II in the singular asymptotic regime is governed by
infinite heteroclinic sequences based on the Kasner
vacuum models}.

The case $\gamma =2$ is exceptional in that there is a
local source, namely a subset of the Jacobs disc (see
Figure I).  We thus conjecture that a typical orbit
with $\gamma =2$ is past asymptotic to an equilibrium
point in  this set. The models described by the
invariant set
$\Sigma_{1} =0$ are also exceptional, since there is a
subset of the Kasner circle $\ck$ that is a local source
(see the end of Section 3).  We thus conjecture that a
typical orbit in this set is past asymptotic to an
equilibrium point in $\ck$.

We conclude this section by comparing our description of
the oscillatory behaviour in the singular regime using
dynamical systems methods in the expansion-normalized
state-space with the descriptions provided by the
Hamiltonian approach (see [18] and [19], pages
63-4) and by the so-called BKL approach ([20], pages
533-8).

For the class of non-tilted SH models of Bianchi types
VIII and IX this oscillatory behaviour is described as a
succession of
\begin{itemize}
\item[i)]
bounces off Bianchi type II potential walls, in the
Hamiltonian approach,

\item[ii)]
changes of the Kasner exponents as described by the BKL
map (see [4], page 236), in the BKL approach, and

\item[iii)]
vacuum Bianchi II orbits (Taub orbits) linking Kasner
equilibrium points in the expansion-normalized
state-space.
\end{itemize}

The presence of tilt leads to a new dynamical phenomenon
which is due to the occurrence of off-diagonal shear
degrees of freedom, and which is described in the above
three approaches as follows:
\begin{itemize}
\item[i)]
bounces off ``centrifugal'' potential walls in the
Hamiltonian approach (see [21], pages 132-4),

\item[ii)] rotation of the Kasner axes, thereby permuting
the Kasner exponents, in the BKL approach (see [22],
pages 640-7), and

\item[iii)]
non-singular Kasner orbits that link physically
equivalent Kasner equilibrium points (see Figure II,
ii) and iii)) in the expansion-normalized state-space.
\end{itemize}

In the expansion-normalized state space, the presence of
tilt leads to an additional new phenomenon, namely
{\it transitions between states with zero tilt and
states with extreme tilt}, described by the non-singular
Kasner orbits that link equilibrium points on $\ck$ and
$\ck_{\rm extreme}$ (see Figure III).  These orbits,
that describe changes in the dynamical significance of
the tilt during the oscillatory regime, do not appear to
have an analogue in the other two approaches.

\section{Discussion}

In this paper we have introduced expansion-normalized
variables to describe the evolution of tilted SH models
of Bianchi type II, and have shown that with this choice
of variable, the Einstein field equations reduce to an
autonomous DE on a five dimensional compact subset of
$\mathbb{R}^{6}$.  Since the state space is compact, the
orbits of the DE have a well-defined asymptotic
behaviour as the dimensionless time $\tau$ tends to $\pm
\infty$.  This property enabled us to give a detailed
description of the dynamics of the above class of models
in the late time regime (see Section 4) and on approach
to the initial singularity (see Section 5).

The description is incomplete in one respect, namely,
that we have not discussed the kinematical properties of
the cosmological fluid.  For a non-tilted SH cosmology
the only non-zero kinematical quantities of the fluid
congruence are the rate of expansion and the rate of
shear, since the acceleration and vorticity are
necessarily zero [12].  On the
other hand, in a tilted SH cosmology, all four
kinematical quantities of the fluid congruence are
non-zero in general [2].  It is
essential to keep in mind, however, that in our
analysis, the dimensionless shear variables \eqref{e2.1}
and the Hubble scalar $H$ describe the kinematics of the
timelike congruence that is normal to the group orbits. 
The kinematical quantities of the fluid congruence can
be expressed in terms of the variables \eqref{e2.1} by
adapting the formulas in  [2] (see sections 1
and 2).  The desired relations, valid for all Bianchi
types, are given in Appendix C, and can easily be
specialized to models of Bianchi type II.  

We can draw the following conclusions from these
relations:

\begin{itemize}

\item[i)]
The cosmological fluid is expanding for all time.  This
fact follows from \eqref{c.1} and \eqref{c.2}, and the
fact that the Hubble scalar of the normal congruence is
assumed to be positive.

\item[ii)]
The vorticity of the fluid is zero\footnote{This result
can be inferred directly from [2] (see
Theorem 3.1).}, but the acceleration is non-zero,
provided that the pressure is non-zero $(\gamma \neq 1)$.

\item[iii)]
If the dimensionless shear of the normal congruence is
small, then the dimensionless shear and acceleration of
the fluid congruence are small.
\end{itemize}

It is known that the evolution of non-tilted SH
cosmologies of Bianchi type II has a very simple
description in the expansion-normalized state space: the
orbit of a model joins two equilibrium
points, or is itself an equilibrium point.\footnote{See
 [4], Section 6.3.2.  We refer to [7] (page
53) and [23], for alternative descriptions.}  Since
equilibrium points correspond to self-similar solutions
of the Einstein field equations, one says that {\it the
models are asymptotically self-similar at the initial
singularity
$(\tau \rightarrow - \infty )$ and at late times} $(\tau \rightarrow
+
\infty )$.  The initial asymptotic state is typically a
Kasner solution and the late time asymptotic state is the
Collins-Stewart perfect fluid solution.  We have seen
that a Bianchi II model can in general accommodate two
tilt degrees of freedom, described by the two
off-diagonal shear variables $\Sigma_{1}$ and
$\Sigma_{3}$.  Our analysis shows that models with only
one tilt degree of freedom (i.e. $\Sigma_{1} = 0$) are
still asymptotically self-similar.  The initial
asymptotic state is again a Kasner state, implying that
the tilt is not dynamically  significant near the
initial singularity.  The late time asymptotic
state\footnote{This asymptotic behaviour has previously
been discussed by [13].} depends on the equation
of state parameter $\gamma$, and is described by the
sequence of bifurcations
$$
F \overset{\gamma = \frac{2}{3}}{\longrightarrow} P II
\overset{\gamma = \frac{10}{7}}{\longrightarrow} P
II_{\rm tilt} ,
$$
a subset of \eqref{e3.2}.  If the second tilt degree of
freedom is activated (i.e. $\Sigma_{1} \neq 0$), we have
seen that evolution is Mixmaster-like near the initial
singularity, and hence the models are no longer
asymptotically self-similar in the singular regime.  The
late time asymptotic state is described by the sequence
of bifurcations \eqref{e3.2}, showing that if $\gamma$
satisfies $\frac{14}{9} <
\gamma < 2$ the tilt becomes extreme.  For models
containing dust $(\gamma = 1)$ or radiation $\left(
\gamma = \frac{4}{3} \right)$, however, the models are
asymptotically self-similar, with the Collins-Stewart
solution as the late time asymptote.

The above results lead to some interesting comparisons
concerning the ways in which various anisotropies, in
particular, {\it spatial curvature, magnetic fields} and
{\it tilt}, affect the dynamics of SH cosmologies.

It is known that the anisotropic spatial curvature in a
SH cosmology, as described by the variables $N_{\alpha}$,
$\alpha = 1,2,3$, affects the stability of the circle of
Kasner equilibrium points (see [4], page 132). If all of
the  $N_{\alpha}$  are non-zero, corresponding to models
of Bianchi types VIII and IX, all Kasner points become
saddles, leading to the existence of infinite
heteroclinic sequences into the past, and hence an
oscillatory singular regime.  It has also been found
that magnetic fields affect the Kasner equilibrium
points  in an analogous way, although the mechanism is
somewhat different in that a magnetic field with more
than one degree of freedom generates off-diagonal shear
components.  We refer to [24]-[27] for details. The
present paper shows that in SH models of Bianchi type
II, the tilt degrees of freedom affect the Kasner circle
in the same way by generating off-diagonal shear
components, leading to an oscillatory singular regime. 
Since Bianchi II orbits are contained in the boundary of
the state space of models of more general Bianchi type
(all except types I and V), we expect that this mechanism
will operate quite generally for tilted Bianchi models.

There are also analogies in the way in which anisotropic
spatial curvature, magnetic fields and tilt affect the
dynamics of SH cosmologies at late times.  If the
equation of state parameter satisfies $0 < \gamma <
\frac{2}{3}$, then all orbits are asymptotic in the
future to the flat FL equilibrium point $F$, implying
that the models isotropize.  Since in this case the
limit of the deceleration parameter $q$ is negative
$( \lim\limits_{\tau \rightarrow - \infty} q = \frac{1}{2} ( 3
\gamma -2) < 0 )$, the models are inflationary. 
For each of the three types of anisotropy, bifurcations
occur as $\gamma$ increases, the result of which is that
these anisotropies influence the dynamics to an
increasing degree at late times.  First, the spatial
curvature destabilizes the flat FL equilibrium point $F$
at the value $\gamma = \frac{2}{3}$, by the creation of
new equilibrium points with non-zero spatial curvature. 
At larger values of $\gamma$,  a magnetic field in
models of Bianchi types II and VI$_{0}$ destabilizes
these new equilibrium points, with the result that the
magnetic field becomes dynamically significant
(see [24]-[25]).  In this paper
we have shown that the tilt in a Bianchi II model acts
in a similar way, becoming dynamically significant if
$\gamma >
\frac{10}{7}$.

There is, however, an important difference between the
effect of tilt in a SH model of Bianchi type II and the
effect of spatial curvature and magnetic fields.  We
have mentioned that the spatial curvature destabilizes
the flat FL point $F$ at $\gamma  = \frac{2}{3}$.  It
has also been shown that a magnetic field leads to a
second destabilization of $F$ at $\gamma = \frac{4}{3}$
(see [26]).  On the other hand, the analysis in
Section 3 shows that in a Bianchi II model the tilt does
not destabilize $F$.  In other words, {\it the flat FL
equilibrium point $F$ is stable with respect to the
off-diagonal shear degrees of freedom that arise in the
presence of tilt}.  In the language of dynamical systems
the tilt increases the dimension of the stable manifold
of the equilibrium point $F$ making it more likely that
orbits will pass close to $F$.  The physical
significance of this result is that {\it in Bianchi II
models tilt increases the probability of intermediate
isotropization with $\Omega
\approx 1$}.  Whether this result is true in general for
tilted SH models requires further investigation.

\section*{Appendix A}

In this Appendix we derive the evolution equations
\eqref{e2.3}-\eqref{e2.4} for the tilted SH models of
Bianchi type II.  We begin by giving the evolution
equations for a general SH model, in terms of
expansion-normalized variables defined relative to the
timelike congruence that is normal to the group orbits. 
These equations follow directly from the standard
orthonormal frame equations for SH models as given in
[4]  (see equation (1.90)-(1.100)), as we now describe.

We introduce a group-invariant orthonormal frame $\{
{\bf e}_{0}, {\bf e}_{\alpha} \}$, where ${\bf e}_{0} =
\bn$ is the unit normal to the group orbits.  The
non-zero commutation functions are ([4], page 39)
$$
\{ H, \sigma_{\alpha \beta}, n_{\alpha \beta},
a_{\alpha}, \Omega_{\alpha} \},
$$
and the energy-momentum tensor
\begin{equation*}
T_{ab} = \mu n_{a} n_{b} + 2 q_{(a} n_{b)} + p (g_{ab} +
n_{a} n_{b}) + \pi_{ab} \tag{A.1} \label{a.1}
\end{equation*}
is described by the source terms
$$
\{ \mu , p, q_{\alpha}, \pi_{\alpha \beta} \},
$$
relative to the chosen orthonormal frame.

The expansion-normalized commutation functions and
source terms are defined by
\begin{alignat*}{5}
\Sigma_{\alpha\beta} &= \frac{\sigma_{\alpha\beta}}{H},
&\quad N_{\alpha\beta} &= \frac{n_{\alpha\beta}}{H},
&\quad A_{\alpha} &= \frac{a_{\alpha}}{H}, &\quad
R_{\alpha} &=\frac{\Omega_{\alpha}}{H},
\tag{A.2}\label{a.2}\\[3mm]
\Omega &= \frac{\mu}{3H^{2}}, &\quad P &=
\frac{p}{3H^{2}}, &\quad Q_{\alpha} &=
\frac{q_{\alpha}}{H^{2}}, &\quad \Pi_{\alpha\beta} &=
\frac{\pi_{\alpha\beta}}{H^{2}}. \tag{A.3} \label{a.3} 
\end{alignat*}

We also require the shear parameter $\Sigma$, defined by
\begin{equation*}
\Sigma^{2} = \frac{\sigma^{2}}{3H^{2}}, \tag{A.4}
\label{a.4}
\end{equation*}
([4] equation (5.17)) and the Hubble-normalized spatial
curvature variables, defined by
\begin{equation*}
\cs_{\alpha\beta} = \frac{^{3}S_{\alpha\beta}}{H^{2}},
\qquad K = - \frac{^{3}R}{6H^{2}}. \tag{A.5} \label{a.5}
\end{equation*}
It follows from \eqref{a.2} and equations (1.28), (1.94)
and (1.95) in [4] that
\begin{equation*}
\Sigma^{2} = \tfrac{1}{6} \Sigma_{\alpha\beta}
\Sigma^{\alpha\beta}, \tag{A.6} \label{a.6}
\end{equation*}
and
\begin{equation*}
\cs_{\alpha\beta} = B_{\alpha\beta} - \tfrac{1}{3}
B_{\mu}^{~\mu} \delta_{\alpha\beta} - 2 \varepsilon^{\mu \nu}
{}_{(\alpha} N_{\beta) \mu} A_{\nu},
\qquad K =
\tfrac{1}{12} B_{\mu}^{~\mu} + A_{\mu} A^{\mu}. \tag{A.7}
\label{a.7}
\end{equation*}
where
$$
B_{\alpha\beta} = 2 N_{\alpha}^{~\mu} N_{\mu \beta} -
N_{\mu}^{~\mu} N_{\alpha\beta}.
$$

We introduce the usual dimensionless time
variable $\tau$ according to
\begin{equation*}
\frac{dt}{d\tau} = \frac{1}{H}, \tag{A.8} \label{a.8}
\end{equation*}
where $t$ is clock time along the normal congruence (see
 [4], page 113). In making the transition to
expansion-normalized variables, the deceleration
parameter $q$ plays an essential role (see [4], pages
112-3).  This dimensionless scalar determines the
evolution of the Hubble scalar $H$ according to 
\begin{equation*}
H' = - (1+q)H, \tag{A.9} \label{a.9}
\end{equation*}
([4], equation (5.11)).  Raychaudhuri's equation ([4],
equation (1.90), in conjunction with \eqref{a.3},
\eqref{a.4}, \eqref{a.8} and \eqref{a.9}, leads to the
following algebraic expression for $q$:
\begin{equation*}
q = 2 \Sigma^{2} + \tfrac{1}{2} (\Omega + 3 P).
\tag{A.10} \label{a.10}
\end{equation*}

The Einstein field equations ([4], equations 
(1.91)-(1.93)), Jacobi
 identities ([4], equations (1.96)-(1.98)) and contracted
Bianchi identities ([4], equations (1.99) and (1.100)),
in conjunction with
\eqref{a.2}, \eqref{a.3}, \eqref{a.5}, \eqref{a.8} and
\eqref{a.9}, now yield the following sets of equations.

\bigskip
\noindent
{\it Gravitational evolution equations:}

\begin{align*}
\Sigma_{\alpha\beta}' &= - (2-q) \Sigma_{\alpha\beta} +
2 \varepsilon^{\mu \nu}_{~~~(\alpha} \Sigma_{\beta)\mu} 
 R_{\nu} - \cs_{\alpha\beta} + \Pi_{\alpha\beta},
\tag{A.11} \label{a.11} \\[3mm]
N_{\alpha\beta}' &= q N_{\alpha\beta} + 2
\Sigma_{(\alpha}^{~~\mu} N_{\beta )\mu} + 2
\varepsilon^{\mu\nu}_{~~~(\alpha} N_{\beta )\mu} 
R_{\nu}, \tag{A.12} \label{a.12} 
\end{align*}
\begin{equation*}
A_{\alpha}' = q A_{\alpha} - \Sigma_{\alpha}^{~\beta}
A_{\beta} + \varepsilon_{\alpha}^{~\mu \nu} A_{\mu} R_{\nu},
\tag{A.13} \label{a.13}
\end{equation*}

\bigskip
\noindent
{\it Source evolution equations:}
\begin{equation*}
\Omega' = (2q-1) \Omega - 3P - \tfrac{1}{3}
\Sigma_{\alpha}^{~\beta} \Pi_{\beta}^{~\alpha} +
\tfrac{2}{3} A_{\alpha} Q^{\alpha} \tag{A.14}
\label{a.14}
\end{equation*}
\begin{equation*}
Q_{\alpha}' = 2 (q-1) Q_{\alpha} -
\Sigma_{\alpha}^{~\beta} Q_{\beta} -
\varepsilon_{\alpha}^{~\mu
\nu} R_{\mu} Q_{\nu} + 3 A^{\beta} \Pi_{\alpha\beta} +
\varepsilon_{\alpha}^{~\mu\nu} N_{\mu}^{~\beta} \Pi_{\beta\nu}.
\tag{A.15} \label{a.15}
\end{equation*}

\bigskip
\noindent
{\it Algebraic equations:}
\begin{equation*}
N_{\alpha}^{~\beta} A_{\beta} = 0, \tag{A.16}
\label{a.16}
\end{equation*}
\begin{equation*}
\Omega = 1 - \Sigma^{2} - K, \tag{A.17} \label{a.17}
\end{equation*}
\begin{equation*}
Q_{\alpha} = 3 \Sigma_{\alpha}^{~\mu} A_{\mu} - 
\varepsilon_{\alpha}^{~\mu \nu}
\Sigma_{\mu}^{~\beta} N_{\beta \nu}. \tag{A.18}
\label{a.18}
\end{equation*}

It should be noted that the source evolution equations
(i.e. the contracted Bianchi identities) are a
consequence of the gravitational evolution equations and
the algebraic equations, and hence contain no additional
information.  It is, however, convenient to use them as
auxiliary equations. 

It should also be noted that equations
\eqref{a.11}-\eqref{a.13}, with \eqref{a.10},
\eqref{a.17} and \eqref{a.18} do not form a fully
determined system of evolution equations.  First, there
is no evolution equation for the variables $R_{\alpha}$
that represent the angular velocity of the spatial
frame $\{ {\bf e}_{\alpha}\}$.  One can in fact use the
freedom in the choice of spatial frame, i.e.\ an
arbitrary time-dependent rotation, to introduce a
non-rotating spatial frame $(R_{\alpha} =0)$.  This
choice is not usually the most convenient one, however. 
In addition, there is neither an evolution equation nor
an algebraic equation for the isotropic pressure $P$ and
the anisotropic stress matrix
$\Pi_{\alpha \beta}$.  Thus {\it in order to obtain a
fully determined system one has to specify the source
and fix the spatial frame}.

We now consider the case of a tilted perfect fluid, with
stress energy tensor
\begin{equation*}
T_{ab} = \tilde{\mu} u_{a} u_{b} + \tilde{p} (g_{ab} +
u_{a} u_{b}), \tag{A.19} \label{a.19}
\end{equation*}
and equation of state
\begin{equation*}
\tilde{p} = (\gamma -1) \tilde{\mu} . \tag{A.20}
\label{a.20}
\end{equation*}
The 4-velocity $\bu$ can be written in the form 
\begin{equation*}
u^{a} = \frac{1}{\sqrt{1-v_{b} v^{b}}} (n^{a} + v^{a}),
\tag{A.21} \label{a.21}
\end{equation*}
where the spacelike vector ${\bf v}$ is orthogonal to the
normal vector $\bn$, and satisfies $0 \leq v_{b} v^{b} <
1$.  The vector ${\bf v}$ is called the {\it tilt vector} of
the fluid.  It has components $(0, v^{\alpha})$ relative
to the orthonormal frame $\{ \bn , {\bf e}_{\alpha} \}$.

We can now express the source terms $P, Q_{\alpha}$ and
$\Pi_{\alpha\beta}$ in terms of the  density
parameter $\Omega$ and the tilt vector $v_{\alpha}$, as
follows.  First, equations \eqref{a.1} and
\eqref{a.19}-\eqref{a.21} imply that
\begin{equation*}
\mu = \frac{G}{1-v^{2}} \tilde{\mu} , \tag{A.22}
\label{a.22}
\end{equation*}
where
\begin{equation*}
G = 1 + ( \gamma -1) v^{2}, \quad v^{2} = v_{\alpha}
v^{\alpha} < 1. \tag{A.23} \label{a.23}
\end{equation*}
By using \eqref{a.22}, in conjunction with \eqref{a.3},
we now obtain
\begin{align*}
P &= \tfrac{1}{3} G^{-1} \left[ 3 (\gamma -1) (1-v^{2})
+ \gamma v^{2} \right] \Omega , \tag{A.24} \label{a.24}\\
Q_{\alpha} &= 3 \gamma G^{-1} \Omega v_{\alpha} ,
\tag{A.25} \label{a.25}\\
\Pi_{\alpha \beta} &= 3 \gamma G^{-1} \Omega
\left( v_{\alpha} v_{\beta} - \tfrac{1}{3} v^{2}
\delta_{\alpha\beta} \right). \tag{A.26} \label{a.26}
\end{align*}

The algebraic equation \eqref{a.18}, with \eqref{a.25},
assumes the form
\begin{equation*}
3 \gamma G^{-1} \Omega v_{\alpha} = 3
\Sigma_{\alpha}^{~\mu} A_{\mu} - \varepsilon_{\alpha}^{~\mu
\nu}
\Sigma_{\mu}^{~\beta} N_{\beta \mu}. \tag{A.27}
\label{a.27}
\end{equation*}
Equations \eqref{a.17} and \eqref{a.27} {\it express the
source terms $\Omega$ and $v_{\alpha}$ algebraically in
terms of the gravitational field variable
$\Sigma_{\alpha \beta}, N_{\alpha \beta}$ and
$A_{\alpha}$}.

It is, however,  convenient to use the source
evolution equations \eqref{a.14} and \eqref{a.15} to
obtain evolution equations for $\Omega$ and
$v_{\alpha}$.  On substituting \eqref{a.24}-\eqref{a.26}
in \eqref{a.14} and \eqref{a.15} and rearranging we
obtain
\begin{equation*}
\Omega' = G^{-1} \left[ 2Gq - (3 \gamma -2) - (2-\gamma
) v^{2} - \gamma \Sigma_{\alpha\beta} v^{\alpha}
v^{\beta} + 2 \gamma A_{\alpha} v^{\alpha} \right]
\Omega , \tag{A.28} \label{a.28}
\end{equation*}
and\footnote{For details of the derivation of this
equation we refer to [28], equations
(65) and (94).}
\begin{equation*}
\begin{split}
v_{\alpha}' &= \frac{v^{\alpha}}{[1- (\gamma -1) v^{2}]}
\bigl[ (3 \gamma - 4) (1-v^{2}) + (2-\gamma )
\Sigma_{\beta \gamma} v^{\beta} v^{\gamma} \\
& \qquad+ \{ (2-\gamma ) - (\gamma -1) (1-v^{2}) \}
A_{\beta} v^{\beta} \bigr] - \Sigma_{\alpha}^{~\beta}
v_{\beta} \\ 
& \qquad\qquad + \varepsilon_{\alpha}^{~\mu\nu} (- R_{\mu} +
N_{\mu}^{~\delta} v_{\delta}) v_{\nu} - v^{2}
A_{\alpha}.
\end{split} \tag{A.29} \label{a.29}
\end{equation*}

Finally, we substitute \eqref{a.24} in \eqref{a.10} to
express $q$ in terms of the gravitational field
variables:
\begin{equation*}
q = 2 \Sigma^{2} + \tfrac{1}{2} G^{-1} \Omega \left[ (3
\gamma -2) (1-v^{2}) + 2 \gamma v^{2} \right] .
\tag{A.30} \label{a.30}
\end{equation*}
We note that \eqref{a.17} can be used to write $q$ in
the form 
\begin{equation}
q = 2 (1-K) - \tfrac{1}{2} G^{-1} \Omega [ 3 (2-\gamma )
(1 - v^{2}) + 2 \gamma v^{2}] \tag{A.31} \label{a.31}
\end{equation}

 For the class of Bianchi II cosmologies
we have $A_{\alpha} = 0$, and in addition we can choose
the spatial frame $\{ {\bf e}_{\alpha} \}$ to be an
eigenframe of the matrix
$N_{\alpha\beta}$, with
\begin{equation}
N_{11} \neq 0, \qquad N_{\alpha\beta} = 0 \quad
\text{otherwise}. \tag{A.32} \label{a.32}
\end{equation}
These restrictions, in conjunction with the constraint
\eqref{a.27}, imply
$$
v_{1} = 0,
$$
assuming $\Omega > 0$ and $\gamma >0$. In addition, 
equation \eqref{a.12} gives
$R_{2} = - \Sigma_{13}$, and $R_{3} = \Sigma_{12}$.  We
are free to perform a rotation in the 23-plane to
get\footnote{This choice is the one made in [2], page
223.}
$$
v_{2} = 0, \quad v_{3} \neq 0.
$$
The constraint \eqref{a.27} now yields
\begin{gather*}
\Sigma_{13} = 0 = R_{2}, \quad \Sigma_{12} \neq 0, \\
3 \gamma \Omega v_{3} = G \Sigma_{12} N_{11} .
\tag{A.33} \label{a.33}
\end{gather*}
Using these results the $\Sigma_{13}'$ equation implies
$$
R_{1} = \Sigma_{23},
$$
and at this stage {\it $R_{\alpha}$ is uniquely
determined in terms of $\Sigma_{\alpha\beta}$}.

We now relabel the variables as follows:
\begin{gather*}
\Sigma_{+} = \tfrac{1}{2} (\Sigma_{22} + \Sigma_{33}),
\qquad \Sigma_{-} = \tfrac{1}{2\sqrt{3}} (\Sigma_{22} -
\Sigma_{33}), \\
\Sigma_{1} = \tfrac{1}{\sqrt{3}} \Sigma_{23}, \qquad
\Sigma_{3} = \tfrac{1}{\sqrt{3}} \Sigma_{12},\\
N_{1} = N_{11} .
\end{gather*}
The set of independent expansion-normalized variables is
$$
(\Sigma_{+}, \Sigma_{-}, \Sigma_{1}, \Sigma_{3}, N_{1},
v_{3}),
$$
subject to one constraint \eqref{a.33}, which we now
write in the form
\begin{equation*}
h(v_{3}) \Omega   =  \Sigma_{3}  N_{1},
\tag{A.34}
\label{a.34}
\end{equation*}
where
\begin{equation*}
h(v_{3}) = \frac{\sqrt{3} \gamma v_{3}}{G}, \tag{A.35}
\label{a.35}
\end{equation*}
 and $G$ is given by
\begin{equation*}
G = 1 + (\gamma -1) v_{3}^{2}, \tag{A.36} \label{a.36}
\end{equation*}
as follows from \eqref{a.23} and the restrictions on
$v_{\alpha}$.  

It should be noted that the off-diagonal shear
components $\Sigma_{1}$ and $\Sigma_{3}$ determine the
angular velocity of the spatial frame according to
\begin{equation*}
(R_{\alpha}) = \sqrt{3} (\Sigma_{1}, 0, \Sigma_{3}).
\tag{A.37} \label{a.37}
\end{equation*}

 The shear parameter has the simple form
\begin{equation*}
\Sigma^{2} = \Sigma_{+}^{2} + \Sigma_{-}^{2} +
\Sigma_{1}^{2} + \Sigma_{3}^{2}, \tag{A.38} \label{a.38}
\end{equation*}
and the curvature parameter $K$ is given by
\begin{equation*}
K = \tfrac{1}{12} N_{1}^{2}, \tag{A.39} \label{a.39}
\end{equation*}
as follows from \eqref{a.7} and \eqref{a.32}.  The
deceleration parameter
$q$, as given by \eqref{a.31}, now assumes the form
\begin{equation*}
q = 2 \left( 1 - \tfrac{1}{12} N_{1}^{2} \right) -
\tfrac{1}{2} G^{-1} \Omega \left[ 3 (2- \gamma )
(1-v_{3}^{2}) + 2 \gamma v_{3}^{2} \right].
\tag{A.40}
\label{a.40}
\end{equation*}
The evolution equations \eqref{e2.3}-\eqref{e2.4} are
now obtained by specializing equations \eqref{a.11},
\eqref{a.12}, \eqref{a.28} and \eqref{a.29} using the
restrictions obtained above.

We conclude this Appendix with an important remark
concerning the interpretation of the expansion-normalized
variables.  In Bianchi II models there are
only two tilt degrees of freedom instead of the
customary three (see [4], Table 9.5, page 211).  The
reason is that the matrix $N_{\alpha\beta}$ has two zero
eigenvalues which, in conjunction with the constraint
\eqref{a.27}, implies that
$$
N_{\alpha}^{~\beta} v_{\beta} = 0,
$$
i.e. the tilt vector lies in the two-dimensional null 
eigenspace of $N_{\alpha}^{~\beta}$.  Our choice of
frame, which leads to $v_{1} = v_{2} = 0$, obscures the
fact that there are two tilt degrees of freedom.  The
constraint \eqref{a.34} shows that {\it the shear
variable $\Sigma_{3}$ represents one of the tilt degrees
of freedom}.  It transpires that {\it the shear variable
$\Sigma_{1}$ represents the second tilt degree of
freedom}, since if $\Sigma_{1} = 0$, the tilt vector is
an eigenvector of $\Sigma_{\alpha\beta}$, i.e.
$$
v_{[\alpha} \Sigma_{\beta ]}^{~\mu} v_{\mu} = 0,
$$
which fixes its direction uniquely.

\section*{Appendix B}

In this Appendix we give some functions that are
monotone along the orbits of the evolution equations,
depending on the value of $\gamma$.

\begin{itemize}
\item[i)]
Consider the function
\begin{equation*}
Z = \frac{\alpha^{3} \Sigma_{1}^{4}
\Sigma_{3}^{2}}{\Omega^{3}} , \tag{B.1} \label{b.1}
\end{equation*}
where
$$
\alpha = \frac{Gv_{3}^{2}}{(1-v_{3}^{2})^{\frac{3}{2} (2
- \gamma)}},
$$
and $G$ is given by equation \eqref{a.36}.  It
follows\footnote{The calculation is simpler if one uses
the constraint (2.2) to express $\Omega$ in terms of the
other variables.} from the evolution equations
\eqref{e2.3}-\eqref{e2.4} that
\begin{equation*}
Z' = 3 (9 \gamma - 14) Z . \tag{B.2} \label{b.2}
\end{equation*}
It thus follows that if $\gamma$ satisfies $0 < \gamma <
2$, $\gamma \neq \frac{14}{9}$, then $Z$ is a monotone
function in the invariant set defined by $\Omega > 0$,
$v_{3} > 0$, $\Sigma_{1} >0$ and $\Sigma_{3} > 0$. 
Furthermore, if $0 < \gamma < \frac{14}{9}$, then
\begin{equation*}
\lim_{\tau \rightarrow + \infty} v_{3} \Sigma_{1} \Sigma_{3} =
0, \quad \lim_{\tau \rightarrow - \infty} (1-v_{3}^{2}) \Omega
= 0, \tag{B.3} \label{b.3}
\end{equation*}
and if $\frac{14}{9} < \gamma <2$, then
\begin{equation*}
\lim_{\tau \rightarrow + \infty} (1-v_{3}^{2}) \Omega = 0,
\quad \lim_{\tau \rightarrow - \infty} v_{3} \Sigma_{1}
\Sigma_{3} = 0. \tag{B.4} \label{b.4}
\end{equation*}

\item[ii)]
For some values of $\gamma$ the tilt variable $v_{3}$ is
monotone.  If $\gamma$ satisfies $0 < \gamma <
\frac{2}{3}$, the factor $3 \gamma - 4 - \Sigma_{+} +
\sqrt{3} \Sigma_{-}$ in the DE for $v_{3}$ is negative
(see \eqref{e2.3}), and hence $v_{3}$ is monotone
decreasing into the future.  It thus follows that if $0
< \gamma < \frac{2}{3}$,
\begin{equation*}
\lim_{\tau \rightarrow + \infty} v_{3} = 0, \quad \lim_{\tau
\rightarrow - \infty} v_{3} =1 . \tag{B.5} \label{b.5}
\end{equation*}
If $\gamma =2$, the DE for $v_{3}$ simplifies to
\begin{equation*}
v_{3}'= (2- \Sigma_{+} + \sqrt{3} \Sigma_{-})
v_{3}.\tag{B.6} \label{b.6}
\end{equation*}
Since $2- \Sigma_{+} + \sqrt{3} \Sigma_{-} > 0$ unless
$(\Sigma_{+}, \Sigma_{-}) = \left( \frac{1}{2}, -
\frac{\sqrt{3}}{2} \right)$, it follows that $v_{3}$ is
monotone increasing into the future.  Thus, if $\gamma
=2$,
\begin{equation*}
\lim_{\tau \rightarrow - \infty} v_{3} = 0. \tag{B.7}
\label{b.7}
\end{equation*}
The DE \eqref{b.6} also implies that if $\gamma =2$, the
extreme tilt set given by $v_{3} =1$ is not an invariant
set, with the result that orbits can pass from the
region $v_{3} < 1$ to the region $v_{3} >1$ in state
space.

\item[iii)]
It follows from \eqref{e2.3} and \eqref{e2.4} that
\begin{equation*}
(\beta \Omega )' = [ 2q - (3 \gamma -2)] (\beta \Omega
), \tag{B.8} \label{b.8}
\end{equation*}
where
$$
\beta = G^{-1} (1 - v_{3}^{2})^{\frac{1}{2} (2-\gamma )}.
$$
The expression  \eqref{a.31} for $q$ can be rearranged
to give
$$
2q - (3 \gamma -2) = 3 (2 - \gamma ) \Sigma^{2} + (2-3
\gamma ) K + \gamma (4-3 \gamma )G^{-1} v^{2} - \Omega,
$$
where $K$ is given by \eqref{a.39}.  Thus, if $\gamma$
satisfies $0 < \gamma < \frac{2}{3}$, $\beta \Omega$ is
monotone increasing, while if $\gamma =2$, $\beta
\Omega$ is monotone decreasing.  It follows that if $0 <
\gamma < \frac{2}{3}$, then
\begin{equation*}
\lim_{\tau \rightarrow + \infty} \Omega =1, \quad \lim_{\tau
\rightarrow -\infty} \Omega = 0. \tag{B.9} \label{b.9}
\end{equation*}

We note that the limits \eqref{b.3}, \eqref{b.4},
\eqref{b.5}, \eqref{b.7} and \eqref{b.9} provide support
for the claims made in Sections 3 and 4 concerning the
asymptotic behaviour.
\end{itemize}

\section*{Appendix C}

In this appendix, we give the equations relating the
kinematical quantities of the fluid congruence in terms
of the dimensionless commutation functions \eqref{a.2}
and \eqref{a.3} associated with the normal congruence. 
The fluid kinematical quantities, namely, the
acceleration $\dot{u}_{a}$, the vorticity vector
$\omega_{a}$ and the rate of shear tensor
$\sigma_{ab}^{\rm fluid}$, are normalized using the
Hubble scalar $H_{\rm fluid}$ of the fluid congruence:
$$
\dot{U}_{a} = \frac{\dot{u}_{a}}{H_{\rm fluid}}, \quad
W_{a} = \frac{\omega_{a}}{H_{\rm fluid}}, \quad
\Sigma_{ab}^{\rm fl} =
\frac{\sigma_{ab}^{\rm fluid}}{H_{\rm fluid}}.
$$
We now give the desired relations, which are obtained by
adapting the results of [2] (see
Sections 1 and 2)\footnote{We thank W.C. Lim for doing
these calculations.}.  Note the relation between our tilt
vector
$v_{a}$ and their vector
$\tilde{c}_{a}$:
$$
v_{a} = (\tanh \beta ) \tilde{c}_{a},
$$
where
$$
\tanh \beta = v.
$$

\bigskip
\noindent
{\it Hubble scalar}:
\begin{equation*}
H_{\rm fluid} = BH, \tag{C.1} \label{c.1}
\end{equation*}
where
\begin{equation*}
B = \cosh \beta \frac{\left[ 1 - \frac{1}{3} (v^{2} +
\Sigma_{\alpha \beta} v^{\alpha} v^{\beta} + 2
A_{\alpha} v^{\alpha}) \right]}{1 - (\gamma -1)v^{2}}.
\tag{C.2} \label{c.2}
\end{equation*}

\bigskip
\noindent
{\it Acceleration}:

$$
\dot{U}_{\alpha} = 3 (\gamma - 1) \cosh \beta \, 
v_{\alpha}, \qquad \dot{U}_{0} = - v^{\alpha}
\dot{U}_{\alpha}.
$$

\bigskip
\noindent
{\it Vorticity}:

$$
W_{\alpha}  = \frac{1}{2B} (N_{\alpha}^{~~\beta}
v_{\beta} + \varepsilon_{\alpha}^{~~\mu \nu} v_{\mu} A_{\nu} +
\cosh^{2} \,  \beta \, N^{\mu \nu} v_{\mu} v_{\nu}
v_{\alpha} ), \quad 
W_{0}  = - v^{\alpha} W_{\alpha}.
$$

\bigskip
\noindent
{\it Shear}:

\begin{align*}
\Sigma_{\alpha \beta}^{f \ell} + \delta_{\alpha \beta}
&= \frac{\cosh \beta}{B} (\Sigma_{\alpha \beta} +
\delta_{\alpha \beta}) - \cosh^{2} \,  \beta\, (4 - 3
\gamma ) v_{\alpha} v_{\beta} \\[2mm]
& \qquad + \frac{\cosh \beta}{B} [N_{(\alpha}^{~~\nu}
\varepsilon_{\beta )\mu\nu} v^{\mu} + A_{(\alpha} v_{\beta )}
- A_{\mu} v^{\mu} \delta_{\alpha \beta} ] , \\[3mm]
\Sigma_{0 \alpha}^{f\ell} &= - \Sigma_{\alpha
\beta}^{f\ell} v^{\beta}, \quad \Sigma_{00}^{f\ell} =
\Sigma_{\alpha \beta}^{f\ell} v^{\alpha} v^{\beta}.
\end{align*}

\section*{References}

\begin{itemize}

\item[1.]
MacCallum, M. A. H. (1979)  {\em General Relativity}
eds. S. W. Hawking and W. Israel (Cambridge University
Press).

\item[2.]
King, A. R. and Ellis, G. F. R. (1973) {\em Commun.
Math. Phys.} {\bf 31} 209-42.

\item[3.]
MacCallum, M. A. H. (1973)  {\em Carg\'{e}se Lectures in
Physics} Volume 6 ed. E Schatzman (Gordon \& Breach).

\item[4.]
Wainwright, J. and Ellis, G. F. R. (1997) {\em Dynamical
Systems in Cosmology} (Cambridge University Press).

\item[5.]
Wainwright, J. (2000)  {\em GRG}, to appear.

\item[6.]
Rosquist, K. and Jantzen, R. T. (1988)  {\em Phys. Rep.}
{\bf 166} 89-124.

\item[7.]
Bogoyavlensky, O. I. (1985) {\em Methods in the
Qualitative Theory of Dynamical Systems in Astrophysics
and Gas Dynamics} Springer-Verlag.

\item[8.]
Peresetsky, A. A. (1985) {\em Topics in Modern
Mathematics} Petroviskii Seminar No.\ 5, ed. O. A.
Oleinik (Consultants Bureau).

\item[9.]
Collins, C. B. and Ellis, G. F. R. (1979)  {\em Phys.
Rep.} {\bf 56} 65-105.

\item[10.]
Hewitt, C. G. and Wainwright, J. (1992)  {\em Phys. Rev.
D} {\bf 46} 4242-52.

\item[11.]
Harnett, D. (1996)  M.Math Thesis, University of
Waterloo.

\item[12.]
Ellis, G. F. R. and MacCallum, M. A. H. (1969)  {\em
Commun. Math. Phys.} {\bf 12} 108-41.

\item[13.]
Hewitt, C. G. (1991) {\em Class. Quantum Grav.} {\bf 8}
L109-14.

\item[14.]
Hsu, L. and Wainwright, J. (1986)  {\em Class Quantum
Grav.} {\bf 3} 1105-24.

\item[15.]
Ma, P. K-H. and Wainwright, J. (1992)  {\em Relativity
Today} ed. Z Perj\'{e}s (Nova Science Publishers);
reprinted in {\em Deterministic Chaos in General
Relativity}, edited by Hobill D, Burd A, Coley A, Plenum
Press, 1994.

\item[16.]
Rendall, A. D. (1997)  {\em Class. Quantum Grav.} {\bf
14} 2341-2356.

\item[17.]
Ringstr\"{o}m, H. (2000)  {\em Class. Quantum Grav.}
{\bf 17} 713-731.

\item[18.]
Misner, C.W. (1969) {\em Phys. Rev. Lett.} {\bf 22},
1071-74.

\item[19.]
Misner, C.W. (1970) in {\em Relativity, Proceedings of
the Relativity Conference in the Midwest}, eds. M.
Carmeli, S.I. Fickler and L. Witten, Plenum Press.

\item[20.]
Belinskii, V. A., Khalatnikov, I. M. and Lifschitz, E. M.
(1970)  {\em Adv. Phys.} {\bf 19}, 525-73.

\item[21.]
Jantzen, R.T. (1987) in {\em Gamow Cosmology}, eds. R.
Ruffini and F. Melchiorri, Proc. Int. School of Physics
E. Fermi, Course LXXXVI, North Holland.

\item[22.]
Belinskii, V. A., Khalatnikov, I. M. and Lifschitz, E. M.
 (1982)  {\em Adv. Phys.} {\bf 31}. 639-67.

\item[23.]
Uggla, C. (1989) {\em Class. Quantum
Grav.} {\bf 6}, 383-96.

\item[24.]
Leblanc, V. G., Kerr, D. and Wainwright, J. (1995)
 {\em Class Quantum Grav.} {\bf 12}, 513-41.

\item[25.]
Leblanc, V. G. (1997) {\em Class. Quantum Grav.} {\bf
14}, 2281-2301.

\item[26.]
Leblanc, V. G. (1998)  {\em Class. Quantum Grav.} {\bf
15}, 1607-26.

\item[27.]
Weaver, M. (2000)  {\em Class. Quantum Grav.} {\bf 17}, 
421-434.

\item[28.]
van Elst, H. and Uggla, C. (1997)  {\em Class. Quantum
Grav.} {\bf 14}, 2673-2695.

\end{itemize}

\end{document}